# Mobile operators as banks or vice-versa?

and: regulators' interest in the best efficiency for payments

L-F Pau, Prof. Mobile Business, Rotterdam School of management
, Prof. Mobile business Copenhagen Business school, and L.M.Ericsson email: lfp.inf@cbs.dk

**Abstract**

This paper addresses the strategic challenges of deposit banks, and payment clearinghouses, posed by the growing role of mobile operators as collectors and payment agents for flows of cash for themselves and third parties. Through analysis and data analysis from selected operators , it is shown that mobile operators achieve as money flow handlers levels of efficiency , profitability ,and risk control comparable with deposit banks – Furthermore , the payment infrastructures deployed by both are found to be quite similar , and  are analyzed in relation to  financial profitability , strategic challenges and opportunities This paves the way to either mobile operators taking a bigger role  ,or for banks to tie up such operators to them even more tightly ,or for alliances/mergers to take place ,all these options being subject to regulatory evolution which is analyzed as well . The consequences are mapped out in operational and regulatory terms -

0.  **Introduction and goals of paper**

The issue of industry convergence , a few say , "has had the attention of both banks and operators for years" ....– Financing issues apart, isn't the in fact that this possible convergence at operational level has been largely ignored, as operators thought that banks  were better at payment services  , while banks thought operators were better at communication services  ? The recent attention given since 2006 to mobile banking by mobile and smartphone terminals makers, leads to the derived strategic question whether these players should sell to operators and end users, or to banks and end users as well (like smart card makers do already). Furthermore , as a convenient explanation , parties implicitly favourable to banks have pushed through the view that mobile networks should be content and transaction neutral , with intelligence and any charging to happen in the province of the end nodes (clearing houses, customers ,and banks ) – While this argument holds for backbone IP or ATM operators , the evolution of service demand/diversity offered by 2.5G and 3G mobile services and terminals goes in the opposite direction for these mobile networks –

Hasn't the above perceptions been reversed by very efficient real time payment , transaction clearing systems and content-on-demand management systems at mobile operators ,but also by some banks analyzing now better and more strategically their information and communication assets from a competitive point of view ? Isn't it time to revisit some roles and regulations to benefit from these efficiencies? Isn't it time for mobile terminal and smartphone makers to position some of their models for early adoption by banks?

The mobile operators become party to financial transactions, but this is only visible where the different regulators so allow! And if the communications regulators might apply rather simple benchmarks to operators' financial flows, banking regulators would not even for payment functions because of the treasury/ cash-flow and money market implications.

These are the issues which this paper , backed by operator data analysis on one hand, and analysis of the payment infrastructures on the other hand , try to raise (or a few might say: revisit ) -

The reader should acknowledge that there is no emphasis on specific Mobile banking (M-Banking) technologies (security, terminals, application software), nor on related market forces from the user demand point of view-

## 1. Mobile banking as a change agent in banks

For those bankers who have experienced and understood to full implications of mobile communication in their business, productivity gains are in a bank or payment clearinghouse, are no longer measured by the substitution of labor force by a tool!

With M-banking , every agent , that is : customer , enterprise and bank personnel alike , becomes a user as well as an information/knowledge source to the service suppliers and to the customers [Anonymous (2003)] – This is, simply put, because so many banking information, knowledge and transaction capabilities resources are brought into the hands of any such agent anytime and anywhere - Thus cooperation and control/audit modes and roles must be redefined between bank customers, banks and third parties – The major difference with the Internet alone as a change agent is the ubiquity and user access, which together cement networking and diffuse changes -

Mobile banking is also a source of value-addition to customers via personalization of features, in both a "push" mode, and in a "pull/definition" role as customers can request or configure some service features, as seen already in Scandinavia with how user define M-banking alerts and refuse some service offerings "pushed" down on them by some banks – Some banks have chosen to add wireless systems as a strategic sales and support channel (e.g. Nordea, BNP Paribas, Bankinter, etc..)

This leads to a basic choice by top bank management: should they promote platforms supporting mobile marketing and consulting, and enhanced mobile services, OR should they allow for and play a role in mobile access to simple e- and M-banking/payment services-The third option, often stated, but not representing a pure business strategy, is to do so called" both; investments and competence build up alone make this very difficult –

That first challenge, leads to a larger one discussed in next section.

## 2. Mobile operators as banks or vice-versa?

This Section raises some issues to be analyzed by cash flow case data analysis, and /or technology analysis in Sections 6, and Sections 2.3, 3, 4. respectively -The reader asking to see issues alongside formal /data analysis should read those Sections as well.

**2.1 : Mobile operators as banks** :With an average 35-65 % (culture and also country dependent) of all mobile generic services being prepaid to the operator over periods of several months to their own offices or via a payment agent (not only banks) ,aren't mobile operators short-term deposit banks holding at any time double digit Billions USD ? Going beyond collection of receivables from their own customers alone, to what extent should operators carry out simple payment processing functions traditionally carried out by banks between their customers and between their customers and third parties? For example, for some mobile operators whose ownership include public utility companies, such third parties could be water, power and cable TV bills –

Furthermore, with mobile operator's capability to handle efficiently and in real-time max. Euro 10-type payments (tickets, parking,..), and their ability in handling bundled service definitions, aren't they micropayment agents [Tewari,O'Mahony (2003)]? In addition, in terms of cross-subsidization, are these micropayment services paid by the generic or value-added transport / communication services?

**2.2: Banks as MVNO's**: When banks "influence" or take over mobile operators via ownership structures, why shouldn't they become mobile virtual operators to capture the operator's client base and their cash transactions covering mobile communication services, but also for other payments enabled via the same transmission and transactions infrastructure?

A third party (bank subsidiary , transactions payment cooperative ,etc ..) can act as an aggregator ,reducing the payment processing and network traffic generated by small-payment users ,but adding this party reduces revenue and fee sharing between the bank and owner of the transmission infrastructure –In a way , mobile roaming operators can be looked upon as actually payment clearing systems , even if historically even the banking shareholders of e.g.. MACH failed to see this.

**2.3: Payment/transactions infrastructures** : Very important is the observation that actually there is not much difference at IT and technology levels, between the customer care and transactions  platforms of mobile operators (see Figure 1 ) ,and those of banks (see Figure 2 ) [Pau , 2003] ! This fact is the result of the evolutions of both layered communications systems architecture, and of banking software systems architecture, in that mobile networks have evolved much faster than fixed networks. The security levels offered by mobile networks inside the infrastructure are also on par with those in banking software, not the least because of added security hardware gives -This means that:

-for a mobile operator to operate also as a payment clearinghouse , is a relatively minor issue , provided the fulfillment systems comply with interbanking data formats , which they even do more and more
-for a bank to operate as a mobile virtual operator using a third party's access networks , is also a relatively minor issue if subscriber data are tagged with bank customer file data , which they even are more and more

## 3. The slow revolution of credit card/new SIM/ Mobile phone combination

It is important not to ignore present day's simple solutions which create M-Banking services for today's networks ,as they provide a pressure to come to grips with the issues raised in the previous two sections –Let us mention as examples, some facts :
-the current use of advanced SIM toolkit (STK) technology (Axalto ,Oberthur , ATOS , Brokat , ActivCard ,etc ) supporting payments , and the SIP Consortium standardizing these features , largely inspired by work at G.I.E Cartes bancaires (France)
-some mobile operators (e.g. Telenor, Orange which have enabled the SIM card as a Visa payment card in selected markets)
-some operators which have enabled a payment credit card as a prepayment card for mobiles (e.g. Wind Italy)
-some payment clearing houses which allow mobile prepaid service reloads from mobile terminals with debit to bank accounts (e.g. Banksys and its new mother company)
-some mobile phones and PDA's have card readers
-the EMPS Electronic Mobile payment services tested by Nordea, VISA Intl, Nokia; it requires the addition of a chip inside terminals which would carry debit, credit, loyalty cars and access codes
-the ability to transmit payment instructions to POS terminals from mobile terminals and PDA's with Bluetooth, without using mobile networks

All these examples, and many more [3] develop a demand for revisiting roles and regulations for banks and mobile operators alike – The difficulty that, in some of the above technical alternatives, stores and POS will have to be reengineered and installed, is largely reduced by the other technical alternatives which bypass such nodes altogether provided the mobile terminals can support them -

## 4. Some Technology usage lessons

Some banks, as well as mobile operators, running M-banking capabilities and services, have learned a few key things, which drive the factors discussed in Sections 1 and 2).
"New technologies" (« technology push »), such as wireless in banking , does not mean only " totally new services" ;most existing services are migrated to multiple access channels ,and enhanced with added functionality when feasible if they mean added value.
Essential limiting factors in M-banking are the « ease-of-use » first (screen size/colour, data entry,..), and technology only second (capabilities of terminals, authentification techniques, ...). Because of:
- the social identification of the mobile terminal with its user;
- the ability to select specific payment transaction types for real-time anywhere downloads;
- the definitely superior capabilities of mobile terminals in terms of personalization and geo-location, bringing the banking branch to the customer,
the « attractiveness » of mobile terminals as personal banking terminals is very strong, and their price/performance/user acceptance is not necessarily second to ATM machines.

Last, the personalization of mobile content and service offerings (the theme: "MyBank is not YourBank"), is ultimately about the only way they both (knowledge and operations) can be provided in a differentiated way by the same banking back office. [McKinsey (2004)] has again recently pointed out that mobile operators that acquire prepaid customers more selectively and offer them well-crafted incentive programs could significantly improve the returns from this segment.

However , as to current or recent global M-payment standards (ETSI, OMA, Mobile Payment Forum, Mobile Payment services association ,M-Commerce Forum, J-Consortium ,etc ..) ,the reality is that , despite claims by some vendors , IT integrators and even operators , we are far from a common standard , so a proprietary "surprise" may still take over ! This means that, alongside open public standards, market based standards controlled by a few parties, may still represent an often incompatible alternative-

Today , if e.g. A UK content provider wants its for-pay offerings to be directly accessible to just the entire UK national mobile subscriber base, he will have to engage individually 5 operators (soon 6) ,  all with different billing systems ! A significant number of retailers will only take plunge into M-Commerce if there is one effective M-payment solution.  As an evolution, far from ideal though, it is necessary to:

 - specify which protocols operators need to use in order to support *different* payment methods (akin PC Internet space with SSL/PKI/Certificates)

 -prioritize amongst the plethora of wireless standards families for M-banking deployment (e.g. drop GSM and deploy from GPRS onwards)

 -integrate the proprietary initiatives of some operators, who cannot wait for the convergence first (e.g. Vodafone: M-Pay service, a pre-pay card for higher value transactions)

-allow and deploy temporarily cross platform solutions with reverse SMS billing (e.g. UK [www.Bango.net](www.Bango.net)); reverse SMS billing includes the capability to make sender (or receiver) validate a monetary transaction they initiate or authorize;

-not to be taken on by some providers who believe that they have "the" solution.

## 5.M-business models implications on payment transactions handled by mobile operators (Personal payments)

For the individual consumer or agent of a product or service (with corresponding fees) other than generic mobile services, the main payment models for individuals are combinations of the following:

1) Deducted from bill: Advertiser pays

2) Deducted from bill: Retailer pays a percentage (comparable to credit cards)

3) Added to bill: Monthly fee added to mobile service bill for access to M-payment services

4) Added to bill: Fee based, similar to SMS charge, per request or transaction

5) Added to or deducted from bill:  Revenue from/to individual from community referrals

6) Added to bill; actual cost of product of service bought by mobile channel

However, in addition, the transmission of the relevant transaction related information warrants transmission costs determined from tariffs. WAP communication service airtime is so expensive that banks find it, sometimes and some places, difficult to charge M-banking fees of type 4) above. Some suppliers and banks wait till operator's tariffs and prices for data transfers (GPRS, EDGE, 3G) have gone down.

In many cases the equation is simple: between credit card or banking fees on one hand, and the total of transmission costs and of the sum of 1) to 5) on the other hand, which is the cheapest channel for the purchase of that product or service? Obviously higher fees have been charged in the past due to "flash or emotional purchases", but with the wireless terminals penetration their use is not reduced to such purchases any more. Even then, total transaction fees (transport, applications, transactions handling) will have to be competitive with e.g. broadband Internet transaction handling fees.

So , just by stating this comparison , which Rotterdam School of management has done [Pau , Vervest(2003)] , a key question is why consumers and enterprises as well, should end paying inflated (or double)  fees on each transaction because mobile operators and payment clearing houses/banks each take their "cut" instead of fulfilling both roles and achieving better economies of scale.

Some operators (e.g. of i-Mode services ,or of some mobile gaming services e.g. in Korea ) , learning from the huge financial success of the old X25 based Minitel services of France Telecom , not only aim at economies of scale , but also at fulfilling transactions for a fee on behalf of third parties such as content owners and banks . At the risk of simplifying, not only do these operators lock margins by service retention at communication level, but they earn additional significant margins by reducing the billing infrastructure investments of third parties, thus taking a percentage of the sum of 1) to 4) and/or billing operator fees (see Figure 3). Total operational margins to operating expenses plus depreciation are above 80 %. The argument is in this sub-section not about analyzing the "success" of these services from the demand and user point of view ,but in terms of benefits to operators adopting new payment models -

## 6. Cases from financial analysis of mobile operators cash flow statements

### 6.1 Objectives and limitations

A sample of publicly listed communications operators with wireless services was taken with spreading over mobile penetration , economic development , countries and currencies , as well as incumbent roles vs. as « pure » plays – The time period covered was from 1998 to 2004 accounting years , although not all data were available for all periods – For incumbents having fixed and other operations as well, mobile operations were taken equal to the ratio of mobile revenues to total revenues – Had to be excluded in a first analysis such operators as Vodafone who do not publish the accounts and data of all the national operators in which they own minority or majority positions-

The sample included : NTT DoCoMo , Singapore Telecom , Orange , Belgacom , Telenor , Nextel Sprint , TDS , Telkom Indonesia , Estonian Telecom . It represents in total more than 70 valid complete annual data sets from the official annual filings with the national securities and exchange commissions-Because of the amount of such data they cannot be reproduced inside this article, but are available to the research community from the annual reports.

## 6.2 Accounting Methodology

The emphasis in the analysis was on Net operating cash flow (NOCF) and its components, and on Free cash flow, together with subscriber, subscription type, employee, CAPEX (capital expenditure for infrastructure and services), and national discount rate data (from the National banks) – In other words, were ignored due in order to analyze intrinsic money flows from operations, cash from from investing and financing-

It is reminded that the Net operating cash flow (NOCF) can be seen in two ways, either as (EBIT: Earnings before interest and tax):
NOCF= EBIT + Depreciation-Tax expenses – Increase in working capital (WCR)
, where the first three terms are the Cash Margin component and the last is the Investment component, or as:
NOCF = Net sales – Cost of good sold –Selling & G&A expenses –Tax expenses – WCR

It is reminded that the free cash flow is:
Free Cash flow = Profit after taxes before interest payment + Depreciation – WCR –New investments

It is also reminded that the Working capital requirements can , from a purely operational point of view , be estimated as including all capital equipment (CAPEX) and staff needed to run the mobile operator infrastructure and back end services -
From a financing perspective, an aggressive strategy occurs when the Cash Margin component is less that Short term debt, and a conservative one is when the reverse happens-

The working assumption is that a mobile operator running a conservative financing strategy and a positive NOCF , should be able for its operations only (ignoring investing activities and financing activities) to operate on zero short term debt and get interest income (at , as an approximation , the prevailing national discount rate) from the NOCF – It is then possible to determine the "NOCF margin" from such interest income from the NOCF , in proportion to total revenues – As NOCF is not made operator segment specific in operator accounts ,it was not possible to determine that "NOCF margin" from mobile operations only , although it is most likely that it is higher than for all the communication services of a mixed service operator- The "NOCF margin" approximates the margin all short term lending/borrowing bank operations would generate in a short term deposit bank.

As it turns out that the shared of postpaid mobile revenues to prepaid mobile revenues across the operators and periods, runs at about 53 %, this situation is comparable to a short term deposit/lending bank where the short term loan portofolio aggregates to 106 % of the short term money deposits –This means that in average the operators have cash operations where the leverage is low and could still be extended-

 "Free cash flow margin" on short term lending/borrowing in a short term deposit bank, after new investments (other than CAPEX and staff operations)-

**6.3 Results and interpretation**

The analysis produces in average over the operators and the years, the following indicators:
-Mobile revenue share: 33, 6 %
-Share postpaid/prepaid in mobile revenues: 55, 13 %
-Free cash flow margin: 1, 94 %
-Net Operating Cash Flow margin: 2,136 %
-Capital expenditures/ NOCF: 84 %

-Annual revenue /employee/year: 341015 Euros

-Free cash flow from operations/employee/year: 110574 Euros

-Cellular subscribers / Mobile operations employee: 1176

-Mobile ARPU (/year): 440 Euros

The main interim conclusion to be drawn is that, in as far as the following indicators are concerned:

-Leverage of short term debt/ short term deposits

-Net operating cash flow margin

-Free cash flow margin

-Free cash flow from operations/employee

-Number of customers / Employee

the average operator in the sample achieves similar financial results from operations than the average short term operations in a bank with limited investment operations, say a postal bank or savings and loans institutions -

## 6.4 Other analyses

In another large research project on Mobile payments and Electronic payment intermediaries in the Netherlands, carried out in collaboration with the European Payment Certification Institute (EPCI) (Waris, F.S.; Maqsoom, F.M.; Pau, L-F. (2006)), it has been shown that, on the basis of the cost and revenue structure for mobile payments in Scandinavia and the Netherlands, mobile operators have a significant financial upside from mobile payments (directly and indirectly through their shareholdings in payment intermediaries). Their margins are also higher than those of electronic payment platforms, largely because of the multiple uses of the same billing platfoms, first for mobile traffic, then for authentification, and last as fee collection platforms on the payment transactions.

## 7. The alternate scenarios

If the opportunities highlighted in Sections 0, 1, 2 are real, existing regulations may prove a hindrance and the regulatory framework may have to evolve.

**7.1. For the regulators** (banking and communications regulators jointly): The regulators have to encourage efficient and secure payments fueling commerce and other services. The non-exclusive alternatives are:

-Banks get individually restricted communications service provider licenses, and lease mobile communications infrastructure
-Bank groups get restricted communications service provider licenses, and lease mobile communications infrastructure
-Mobile operators (genuine or virtual) get additional deposit bank licenses on demand
-Mobile operators (genuine or virtual) gets automatically deposit bank licenses as part of their communications license; this option is of great appeal to developing countries where the banking infrastructure, coverage and trust are far lower than those of mobile operators [Centeno (2003)]
-Mobile payment services are authorized to be opened up for licensing by third parties for their own customers (oil companies, physical transport networks, health system,..)

In any case, communications and banking regulations would have to share mandatory prescriptions in terms of cash and short term debt, which would often be a change from the present situation.

7.2: **For a Bank**: The alternatives are to choose amongst:

1) Existing bank card system operator(s) own and manage servers, with proprietary applications, to handle multiple channels such as mobile; a leased line/IP access solutions to the GGSN node of a mobile operator is sufficient; there is the option for a bank of owning an SMS/MMS Service center
2) Banks outsource some channels (such as mobile) to IT service companies if accepted by operators and not too expensive, and obey IT industry standards
3) Banks internationally create, or cooperate with, existing third party service suppliers to several mobile operators (e.g. roaming/ authentification suppliers) to enhance their services to transactions .The bank than would align it with communications industry standards
It should be noted also that the issue spills over to payment terminals standardization; Captin, a project aiming in a first phase to standardise payment terminal to acquirer protocols (www.captin.org) is very much aware that M-payments will face identical problems of interoperability, unless players address the issues 1)-3) above from the start and create a sufficient volume of users / subscribers using one interoperable standard.

**7.3: For a mobile operator:** The non-exclusive alternatives are to:
-1)Delegate, for a % of the transaction fees (volume based) , fulfilment ,collection and risk management ,to banks or banking payment cooperatives ;this is the by default most frequent currently found option
2)Own , alone or jointly , payment clearinghouse , bank(s) or consumer credit companies to perform the services listed under 1) ;this is also quite common today, although different bank/credit card consortia offer competing vehicles
3) Apply for a deposit banking license in their own name, manage risks and reinsurance, and handle collection on behalf of third parties (content owners, administrations and public services)
4) Split between 1) for large transactions, and 3) for small transactions and reloads (for mobile services as well content)

In all cases the mobile operators must:
-develop in ITU, ETSI, 3GPP, IETF, OMG, and OMA open public standards supporting payment technologies and their end-to-end security and tracing
- develop media campaigns to « prove » the equal/ higher security of mobile services, than many bank card solutions
- not use proprietary technologies, on interoperable networks and terminals.... This is doomed to fail in terms of deployment, adaptations and costs!
 Furthermore , the stance taken by the communications and/or banking regulators, might imply that operators would have to maintain cash positions they are not used to in many cases (due to their usual high debt leverage).

7.4. **For a manufacturer of mobile terminals and smartphones:** The main implications are whether to consider banks and cash payment operators (like parking lots, pharmacies, etc) as sales channels supplementing mobile operators and direct/indirect sales to end users. Some banks maintain expensive services and distribution of token cards to be used in connection with Internet banking or secure payment applications. By enabling at design stage banking systems compliance and key generation, the mobile terminals and smartphone makers could offer the banks better devices, services, CRM, and continuous communications charges flows to banks.

**CONCLUSION**

Many of the central issues raised in this paper illustrate the often difficult recognition by some parties of the co-existence of two basic models in M-Business:
- The centralized model in M-Business: where trade, transaction rules and some generic business processes, are embedded right inside the core of enterprise and public communication networks, managed by these two parties, with a flow of service fees
-The decentralized model in M-Business , with personalized terminals and services , offering full mobility and capability offerings , managed by user-driven exploration matching algorithms, and billing along the value chain ; in this model too users are also information and know-how producers
Obviously the banks come from the centralized model for their mass operations, but aim for the second model as value creators –The mobile operators too have the same profile, while coping with difficulties from the decentralized model they helped propagate – Thus indeed, mobile operators and some banks should be allowed to "converge" as mobility based IT slowly penetrates the conventional IT backbones of banks ,and as the efficiency of transactions handling by mobile operators can give advantages to the banks adopting them while influencing the value added communication services they provide .

**Figure 1**: Mobile communications systems customer management and billing architecture

**Figure 2**: Payment/banking systems customer management and settlement architecture

**Figure 3:** Collection of content owners' receivables and of own revenues by a mobile operator (e.g. i-Mode, Minitel)

# Banking systems / Communications systems

| Billing (& hot..) | Cash flow mgt | Infrastructure | OSS/NM |

**(Operator) Enterprise integration services**

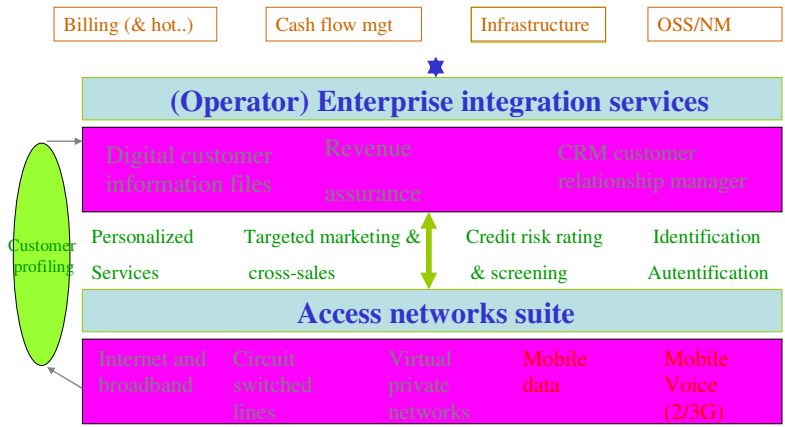

Digital customer information files | Revenue assurance | CRM customer relationship manager

Customer profiling

Personalized Services | Targeted marketing & cross-sales | Credit risk rating & screening | Identification Autentification

**Access networks suite**

Internet and broadband | Circuit switched lines | Virtual private networks | Mobile data | Mobile Voice (2/3G)

# Banking systems / Communications systems

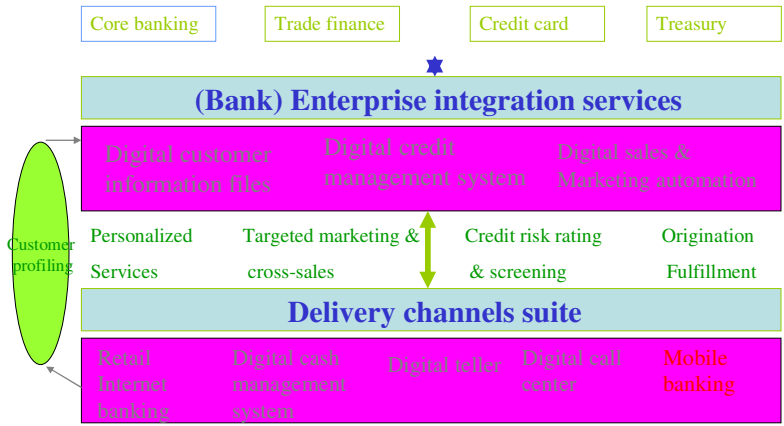

# Content provider bill collection system (e.g. Minitel/iMode) (very many content channels)

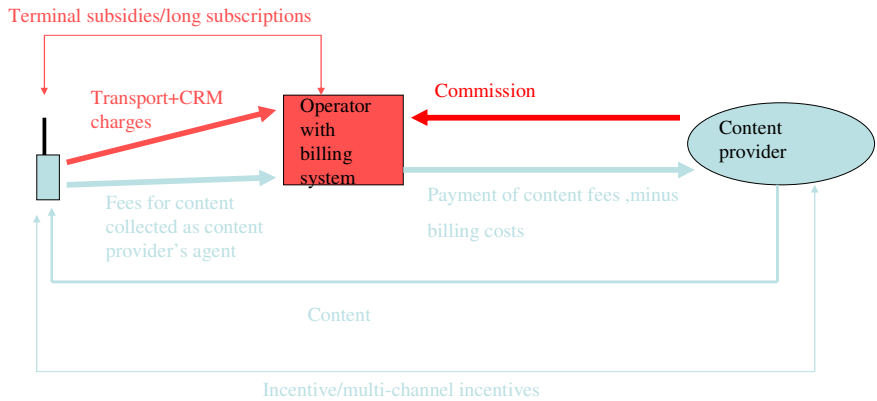